
\input harvmac
\def\Titleline#1{\vskip -0.4in \centerline{\titlefont #1}
\abstractfont\vskip 0.5in}
\def\npb#1#2#3{{\it Nucl.\ Phys.} {\bf B#1} (19#2) #3}
\def\plb#1#2#3{{\it Phys.\ Lett.} {\bf #1B} (19#2) #3}

\def\prd#1#2#3{{\it Phys.\ Rev.} {\bf D#1} (19#2) #3}
\def\pr0#1#2#3{{\it Phys.\ Rev.} {\bf #1} (19#2) #3}
\def\p{\partial}

\def\frac#1#2{{#1 \over #2}}

\def\ie{{\it i.e.}}
\def\sltwo{SL(2,{\bf R})}
\def\AB{\overline{A}}
\def\pb{\overline{\p}}
\def\rM{\frac{r_M}{2}}
\def\rE{\frac{r_E}{2}}
\def\thL{\frac{\theta_L}{2}}
\def\thR{\frac{\theta_R}{2}}
\def\Tr{{\rm Tr}\;}
\def\const{{\rm const}}
\def\intz{\int d^2z\;}
\def\phih{\frac{\phi}{2}}
\def\ln{{\rm ln}\;}
\def\petraddress{\vskip.1in \centerline{Enrico Fermi Institute}
\centerline{University of Chicago}
\centerline{5640 South Ellis Avenue} \centerline{Chicago, IL 60637, USA}
\centerline{e-mail: horava@yukawa.uchicago.edu}}
%
\Title{EFI-91-57}{\ }
\Titleline{Some Exact Solutions of String Theory}
\Titleline{in Four and Five Dimensions}
%
\centerline{Petr Ho\v{r}ava\footnote{$^{\ast}$}{Robert R. McCormick
Fellow; work also supported by
the NSF under Grant No.\ PHY90-00386;
the DOE under Grant No.\ DEFG02-90ER40560;
the Czechoslovak Chart 77 Foundation;
and the \v{C}SAV under Grant No.\ 91-11045.}}
\petraddress
%
\vskip .4in
We find several classes of exact classical solutions of critical bosonic
string theory, constructed as twisted products of one Euclidean and one
Minkowskian 2D black hole coset.  One class of these solutions leads
(after tensoring with free scalars and supersymmetrizing) to a rotating
version of the recently discovered exact black fivebrane.
Another class represents a one-parameter family of axisymmetric stationary
four-dimensional targets with horizons.  Global properties and target
duality of the 4D solutions are briefly analyzed.

\Date{October 1991}
%
%
%
\newsec{Introduction}

Exact solutions of (perturbative) string theory that do not represent
strings in static spacetimes with flat time coordinate
(plus possible internal degrees of freedom), but rather describe
spacetimes of nontrivial metric properties,
have attracted much interest recently.  One of the most excellent
examples are Witten's black holes in 2D string theory
\ref\witten{E. Witten, \prd{44}{91}{314}}, as well as their
generalizations to charged black holes in 2D \ref\ishib{N. Ishibashi,
M. Li and A.R. Steif, `Two Dimensional Charged Black Holes in String
Theory,' Santa Barbara preprint UCSBTH-91-28 (June 1991)}, black
strings in 3D \ref\horhor{J.H. Horne and G.T. Horowitz, `Exact Black
String Solutions in Three Dimensions,' Santa Barbara preprint
UCSBTH-91-39 (July 1991)}, and exact fivebranes in 10D superstring theory
\ref\gidstro{S.B. Giddings and A. Strominger, `Exact Black Fivebranes
in Critical Superstring Theory,' Santa Barbara preprint UCSBTH-91-35
(July 1991)}.  With these solutions at hand, one naturally wonders
whether the techniques can be used to construct exact solutions
of string theory in four dimensions.

To answer this question in the affirmative, let us start with Witten's
black hole coset in 2D spacetime with Minkowskian signature.  We might
obtain a model which describes strings in a 4D manifold with Minkowskian
signature by tensoring the 2D black hole with another conformal
field theory, describing strings in a 2D manifold of Euclidean signature.
Obviously, we have one excellent candidate for such a manifold:  The 2D
black hole itself, now in the Euclidean regime.

To get a nontrivial 4D spacetime, we would like to allow a `twist' in
the product of the two conformal field theories.  Technically, the basic
idea of this paper is to start with a direct product of two WZW models,
and gauge a group that acts nontrivially on both of them, thus producing
a conformal field theory that is no longer a direct product
(compare \horhor ).  In the case
of two 2D black holes, a naive way of doing so might be as follows.
Starting with the direct product of two $\sltwo$ WZW models
(referred to as $\sltwo_{M,E}$ throughout),
\eqn\dirsl{\eqalign{\CL _{\rm WZW}=&\;\frac{ik_M}{4\pi}\intz \Tr
(g_M^{-1}\p g_M\; g_M^{-1}\pb g_M)-ik_M\Gamma(g_M)\cr
&+\frac{ik_E}{4\pi}\intz \Tr
(g_E^{-1}\p g_E\; g_E^{-1}\pb g_E)-ik_E\Gamma(g_E)\cr}}
where $g_{M,E}\in \sltwo _{M,E}$, we will gauge two Abelian symmetry
groups of the model.  First, we will gauge the compact Abelian group
\eqn\gacomp{g_E\rightarrow h_E\; g_E\; h_E}
with $h_E$
generated by $\pmatrix{0&1\cr -1&0\cr}$.  At this stage, we get a direct
product of the 2D black hole coset with Euclidean signature in sector $E$
and the ungauged WZW model in sector $M$.  In the second step, we will
gauge the noncompact group
\eqn\ganonc{g_M\rightarrow h_M\; g_M\; h_M, \qquad g_E\rightarrow
h_E^{\alpha}\; g_E\; h_E^{-\alpha}}
with $h_M$ generated by $\pmatrix{1&0\cr 0&-1}$.  Here
$\alpha\in {\bf R}$ is a `distortion parameter':  For $\alpha=0$, we get
a direct product of one Euclidean and one Minkowskian 2D black hole.

The total conformal anomaly of the gauged model is
\eqn\cano{c=\frac{3k_M}{k_M-2}+\frac{3k_E}{k_E-2}-2.}
To get a critical theory in four dimensions, we set $c=26$.  This condition
restricts the values of $k_{M,E}$ to
\eqn\kvalue{k_{M,E}=k\pm \tilde{k},}
with
\eqn\kval{\tilde{k}=\pm\sqrt{(k-\frac{28}{11})(k-2)}.}
We will further restrict ourselves to
\eqn\resk{k\geq \frac{28}{11},}
to avoid complex values of the levels.%
\foot{We have omitted here the other region with real $k_{M,E}$, namely
$k\leq \frac{8}{5}$.  This region would correspond to the
analytic continuation of one of the $\sltwo$'s to $SU(2)$ (see below).}

Let us parametrize the group manifolds by their Euler angles:
\eqn\euangm{\eqalign{g_M=&\pmatrix{e^{t_L/2}&0\cr 0&e^{-t_L/2}\cr}
\pmatrix{\cosh \rM &\sinh \rM \cr \sinh \rM &\cosh \rM\cr}
\pmatrix{e^{-t_R/2}&0\cr 0&e^{t_R/2}\cr},\cr
g_E=&\pmatrix{\cos \thL &\sin \thL \cr -\sin \thL &\cos \thL\cr}
\pmatrix{\cosh \rE &\sinh \rE \cr \sinh \rE &\cosh \rE\cr}
\pmatrix{\cos \thR &-\sin \thR \cr \sin \thR &\cos \thR\cr},}}
with $r_M\in {\bf R}, r_E\in [0,\infty), t_{L,R}\in {\bf R},$ and
$\theta_{L,R}\in [0,2\pi ]$,%
\foot{Two facts seem worth stressing.  First, the Euler angle parametrization
of $g_M$ we have used does not cover the whole $\sltwo _M$ manifold.
Second, the ranges for $\theta_{L,R},r_E$ cover $SO(2,1)$ rather than
its double cover $\sltwo$.}
and denote the gauge fields that correspond to \gacomp\ and \ganonc\ by
$A_E,\AB_E$ and $A_M,\AB_M$ respectively.  Upon choosing a unitary gauge
by setting
\eqn\ungau{t_L=t_R\equiv t,\qquad \theta_L=\theta_R\equiv \theta,}
we arrive at the following Lagrangian:
\eqn\lagcoset{\eqalign{\CL_{\rm 4D}=&\;\CL_{\rm WZW}+\frac{ik_M}{\pi}
\intz \sinh ^2\rM (\AB_M\p t-A_M\pb t)\cr
&\qquad \qquad \qquad \qquad +\frac{2ik_M}{\pi}\intz
\cosh ^2\rM \; A_M\AB_M\cr
&+\frac{ik_E}{\pi}\intz \sinh ^2\rE \; (A_E\pb \theta
-\AB_E\p \theta +\alpha A_M\pb \theta +\alpha\AB_M\p\theta )\cr
&+\frac{2ik_E}{\pi}\intz (\alpha ^2\sinh ^2\rE \; A_M\AB_M-
\cosh ^2\rE \; A_E\AB_E)\cr
&+\frac{ik_E}{\pi}\intz \alpha (A_E\AB _M-A_M\AB _E)(\cosh ^2\rE +
\sinh ^2\rE ).}}
As the gauge fields enter quadratically, they can be integrated out
by solving their equations of motion.  The final (lowest order)
Lagrangian reads
\eqn\lagsigma{\eqalign{\CL_{\rm 4D}=&\;\frac{i}{2\pi}\intz
\left( \frac{k_E}{4}\p r_E\pb r_E
+\frac{k_E\sinh ^2\rM(\cosh ^2\rE -K)}{\cosh ^2\rM \cosh ^2\rE -K}
\p\theta\pb\theta\right.  \cr
&\qquad\qquad +\frac{k_M}{4}\p r_M\pb r_M -k_M\frac{\sinh ^2\rE
( \cosh ^2\rM -K)}{\cosh ^2\rM \cosh ^2\rE
-K}\p t\pb t\cr
&\left.+\frac{\alpha k_E\sinh ^2\rM \sinh ^2\rE}{2(
\cosh ^2\rM \cosh ^2\rE -K)}(\p\theta\pb t-\pb\theta\p t)
\right) ,\cr}}
with $K$ a shorthand for $4\alpha ^2k_E/k_M$.
Obviously, it describes strings in a 4D target, with the following
metric and antisymmetric tensor background:
\eqn\backsigma{\eqalign{ds^2=&\; \frac{k_E}{4}dr_E^2+\frac{k_M}{4}dr_M^2+
k_E\frac{\sinh ^2\rM(\cosh ^2\rE -K)}{\cosh ^2\rM \cosh ^2\rE -K}
d\theta ^2\cr &\qquad \qquad \qquad \qquad
-k_M\frac{\sinh ^2\rE ( \cosh ^2\rM -K)}{\cosh ^2\rM \cosh ^2\rE
-K}dt^2 ,\cr
B=&\;\frac{\alpha k_E\sinh ^2\rM \sinh ^2\rE}{2(\cosh ^2\rM \cosh ^2\rE -K)}
d\theta\wedge dt.\cr}}
The nontrivial determinant
coming from the integration over the gauge fields leads to a dilaton
background,
\eqn\dilone{\Phi =\ln (\cosh ^2\rM \cosh ^2\rE -K)+\const .}
Throughout the paper, we will only consider sigma-model metrics.  The
so-called canonical metrics can be obtained from the sigma-model metrics
by a proper rescaling by an exponential of $\Phi$.

A priori we might have expected that the 4D theory would have an
exact $U(1)\times U(1)$ symmetry.  Indeed, two abelian
symmetries survive our gauging of the WZW model.  Without any additional
arguments, our conformal field theory seems to represent a class of
exact stationary and axisymmetric solutions of string theory in four
dimensions.  Unfortunately, this is not true, the reason being that the
gauging we have attempted to do is in fact anomalous.
Indeed \ref\witan{E. Witten, `On
Holomorphic Factorization of WZW and Coset Models,' IAS preprint
IASSNS-HEP-91/25 (June 1991)}, we can easily check that the
anomaly-cancellation condition,
\eqn\ancanc{\Tr (T_{a,L}T_{b,L}-T_{a,R}T_{b,R})=0,}
is not met.  (Here we have used the notation of \witan , \ie\ $a,b$ are
gauge group indices and $T_{a,L},T_{a,R}$ generate the gauge group action
on the WZW fields, $\delta g=\epsilon ^a\{ T_{a,L}\cdot g+g\cdot
T_{a,R}\} $.)  We thus cannot hope that \backsigma\ is more than
a solution of the low-energy effective approximation to string theory.

To obtain genuine solutions of string theory, we will modify our basic
strategy in two directions.  First, we will obtain in section 2 a class of
models in five dimensions, simply by gauging the $U(1)$ group that acts
on both of the $\sltwo$'s, and forgetting about the other $U(1)$.
As we will see,
this leads (after continuing analytically, tensoring with free bosons, and
supersymmetrizing) to rotating versions of the recently discovered
\ref\chs{C.G. Callan, J.A. Harvey and A. Strominger,
\npb{359}{91}{611}}\gidstro\ fivebrane solitons of superstring theory.
Second, in section 3 we will modify the action of
$U(1)\times U(1)$ so as to avoid the violation of Witten's condition
of non-anomalousness.  This will result in a class of
exact, stationary and axisymmetric solutions in four dimensions.

\newsec{Exact Rotating Black Fivebranes from 2D Black Holes}

Let us now start with the tensor product Lagrangian \dirsl , and gauge
the $U(1)$ group acting by
\eqn\fiveact{g_E\rightarrow h_E\; g_E\; h_E,\qquad g_M\rightarrow
h_M^{\beta}\; g_M\; h_M^{\beta}}
with $\beta$ a distortion parameter.
Upon gauging this group, the Lagrangian becomes
\eqn\figala{\eqalign{\CL =&\;\CL _{\rm WZW}+
\frac{ik_E}{2\pi}\intz A\;\Tr\!
\left(\!\pmatrix{0&1\cr -1&0\cr}\pb g_Eg_E^{-1}\right)\cr
&+\frac{ik_E}{2\pi}\intz \AB\;\Tr\!
\left(\!\pmatrix{0&1\cr -1&0\cr}g_E^{-1}\p g_E\right)\cr
&+\frac{ik_E}{2\pi}\intz A\AB \left[ -2+
\Tr\!\left(\!\pmatrix{0&1\cr -1&0\cr}g_E\pmatrix{0&1\cr -1&0\cr}
g_E^{-1}\right)\right]\cr
&+\frac{ik_M}{2\pi}\intz \beta A\;\Tr\!
\left(\!\pmatrix{1&0\cr 0&-1\cr}\pb g_Mg_M^{-1}\right)\cr
&+\frac{ik_M}{2\pi}\intz \beta\AB\;\Tr\!
\left(\!\pmatrix{1&0\cr 0&-1\cr}g_M^{-1}\p g_M\right)\cr
&+\frac{ik_M}{2\pi}\intz \beta ^2A\AB \left[ 2+
\Tr\!\left(\!\pmatrix{1&0\cr 0&-1\cr}g_M\pmatrix{1&0\cr 0&-1\cr}
g_M^{-1}\right)\right]\cr}}
where we have denoted by $A, \AB$ the gauge field associated with \fiveact .
Upon parametrizing the group manifolds as in \euangm\ and fixing the
gauge by
\eqn\gffi{\theta_L=\theta_R\equiv\theta ,}
we observe that the model describes the following five-dimensional
background (to lowest order):
\eqn\bacfive{\eqalign{ds_{\rm 5D}^2=&\;\frac{k_M}{4}dr_M^2+
\frac{k_E}{4}dr_E^2+
\frac{k_E\sinh ^2\rE (1-L\cosh ^2\rM )}{\cosh ^2\rE -L\cosh ^2\rM}
d\theta ^2\cr
&+\frac{k_M\sinh ^2\rM (L-\cosh ^2\rE )}{\cosh ^2\rE -L\cosh ^2\rM}
dt^2 +\frac{2k_M\beta\sinh ^2\rE \sinh ^2\rM}{\cosh ^2\rE -L\cosh ^2\rM}
dt\; d\theta\cr
&\qquad\qquad +\frac{k_M\cosh ^2\rM\cosh ^2\rE}{\cosh ^2\rE
-L\cosh ^2\rM}d\tilde{t}^2,\cr
B=&\;\frac{k_M\sinh ^2\rM \cosh ^2\rE}{\cosh ^2\rE -L\cosh ^2\rM}dt\wedge
d\tilde{t}-\frac{\beta k_M\sinh ^2\rE\cosh ^2\rM}{\cosh ^2\rE-L\cosh ^2\rM}
d\theta\wedge d\tilde{t} ,\cr
\Phi=&\;\ln (\cosh ^2\rE -L\cosh ^2\rM )+\const\cr}}
where $t,\tilde{t}=\half (t_L\pm t_R)$ and $L=\beta ^2k_M/k_E$.  As \gffi\
does not fix the gauge completely, $\tilde{t}$ is orbifoldized, and
$\tilde{t}\equiv\tilde{t}+2\pi\beta$.

While this geometry is interesting in itself, we can find connections to some
results obtained recently \chs ,\gidstro ,\ref\garf{D. Garfinkle, G. Horowitz
and A. Strominger, \prd{43}{91}{3140}}--\nref\gibbons{G.W. Gibbons and
K. Maeda, \npb{298}{88}{741}}\nref\horstr{G.T. Horowitz and A. Strominger,
\npb{360}{91}{197}}\nref\callrev{C.G. Callan, Jr., `Instantons and Solitons
in Heterotic String Theory,' Princeton U. preprint PUPT-1278 (June 1991)}
\ref\khuri{R.R. Khuri, \plb{259}{91}{261}} by continuing it analytically
to a gauged $\sltwo\times SU(2)$ WZW model.  Upon parametrizing the $SU(2)$
group manifold by its Euler angles,
\eqn\eusutwo{g=\pmatrix{e^{i\theta_L/2}&0\cr 0&e^{-i\theta_L/2}\cr}
\pmatrix{\cos \frac{\phi}{2}&i\sin \frac{\phi}{2}\cr i\sin\frac{\phi}{2}&
\cos\frac{\phi}{2}\cr}\pmatrix{e^{-i\theta_R/2}&0\cr 0&e^{i\theta_R/2}\cr},}
with $\phi\in [0,\pi )$ and the ranges for $\theta_{L,R}$ as before,
we can see that the corresponding gauged model is related to the one
constructed previously, by the analytic continuation of $r_E$ to $\phi$ via
$r_E=i\phi$.
In addition, this analytic continuation has to be supplemented with the
sign reversal of the level, in order to preserve the relative metric signature
of the two group manifolds.  We will thus assume $k\equiv -k_E \geq 0$
henceforth, as well as reverse the sign of $L$ so as to ensure $L\geq 0$.
The central charge of the model is
\eqn\centnew{c=\frac{3k_M}{k_M-2}+\frac{3k}{k+2}-1}
and $k$ is restricted by unitarity to a discrete set
of values, as usual.  (We don't set $c=26$ here, as it is more interesting
to tensor the model with five free scalars and supersymmetrize it.  Imposing
$c_{\rm tot}=15$ afterwards, we obtain a solution of $N=1$ superstring
theory.)
With these conventions, we get
\eqn\fivebrane{\eqalign{ds_{\rm 5D}^2=&\;\frac{k_M}{4}dr_M^2+
\frac{k}{4}d\phi ^2+
\frac{k\sin ^2\phih (1+L\cosh ^2\rM )}{\cos ^2\phih +L\cosh ^2\rM}
d\theta ^2\cr
&-\frac{k_M\sinh ^2\rM (L+\cos ^2\phih )}{\cos ^2\phih +L\cosh ^2\rM}
dt^2 -\frac{2k_M\beta\sin ^2\phih \sinh ^2\rM}{\cos ^2\phih +L\cosh ^2\rM}
dt\; d\theta\cr
&\qquad\qquad +\frac{k_M\cosh ^2\rM\cos ^2\phih}{\cos ^2\phih
+L\cosh ^2\rM}d\tilde{t}^2,\cr
B=&\;\frac{k_M\sinh ^2\rM\cos ^2\phih}{\cos ^2\phih +L\cosh ^2\rM}dt\wedge
d\tilde{t}+\frac{\beta k_M\sin ^2\phih \cosh ^2\rM}{\cos ^2\phih +L\cosh ^2
\rM}d\theta\wedge d\tilde{t} ,\cr
\Phi=&\;\ln (\cos ^2\phih +L\cosh ^2\rM )+\const .\cr}}
This model is a rotating analog of the fivebrane
soliton discovered in \chs\gidstro .  The core of the fivebrane is
surrounded by a spherical horizon%
\foot{Our coordinates $\phi ,\theta ,\tilde{t}$ do parametrize a 3-sphere,
albeit in an unusual manner.  More standard coordinates on the sphere
would result from an alternative gauge choice in the gauged WZW model,
namely $t_L=t_R\equiv 0$.}
localized at $r_M=0$.  At fixed $t$, the
horizon inherits the following metric:
\eqn\hormemb{ds_{\rm horizon}^2=\frac{k}{4}d\phi ^2+
\frac{k\sin ^2\phih (1+L)}{\cos ^2\phih +L}
d\theta ^2+\frac{k_M\cos ^2\phih}{\cos ^2\phih
+L}d\tilde{t}^2.}
The exact black fivebrane of \gidstro\ has the structure of the direct product
of $SU(2)$ and $\sltwo /U(1)$, which we recover in the limit of $L\rightarrow
\infty$.

It is worth noting that the model is indeed not asymptotically flat,
rather it is asymptotic to $S^3\times {\bf R}$, as can be seen
in the $r_M\rightarrow \infty$ limit of the metric:%
\foot{Actually, one might expect the model
to represent a limiting, exactly solvable case of a class of solutions
to the low-energy action of string theory, quite analogously as in \chs .
These low-energy solutions can be expected to open the throat at
infinity.  I am indebted to Jeff Harvey for illuminating discussions
on this point.}
\eqn\rlimas{\eqalign{ds_{\rm 5D}^2\ \rightarrow&\ \frac{k_M}{4}dr_M^2+
\frac{k}{4}d\phi ^2+k\sin ^2\phih d\theta ^2
-(k_M+\frac{k}{\beta ^2}\cos ^2\phih )dt^2\cr
&\qquad\qquad -\frac{2k}{\beta}\sin ^2\phih dt\; d\theta
+\frac{k}{\beta ^2}\cos ^2\phih d\tilde{t}^2.\cr}}
We can also observe dragging of inertial frames, a typical effect
of rotating bodies in general relativity.

As we have argued that our class of conformal field theories represents
essentially
a rotating deformation of the fivebrane solution constructed by Giddings
and Strominger, it is natural to look for the exact marginal vertex
operator that governs this deformation.  In the approximation of
\fivebrane , the vertex operator can be easily identified as
\eqn\marg{V_{\rm marg}\sim\frac{ik}{2\pi}\intz \sin ^2\phih \left[
\tanh ^2\rM (\p t\pb\theta +\pb t\p\theta )-(\p\theta\pb \tilde{t}-
\pb\theta\p\tilde{t})\right].}
This indeed represents a lowest order approximation to an exactly
margninal operator.  The exact form of the operator can be identified
by looking at the full-fledged coset Lagrangian, leading to
\eqn\margex{V_{\rm marg}=\frac{ik}{2\pi}\intz \left[ A\;\Tr\!
\left(\!\pmatrix{0&1\cr -1&0\cr}\pb g_Eg_E^{-1}\right) +\AB\;\Tr\!
\left(\!\pmatrix{0&1\cr -1&0\cr}g_E^{-1}\p g_E\right)\right].}
One can easily check that after integrating out the gauge field in \margex ,
one arrives at \marg .  Note the interesting fact that the exactly
marginal vertex operator \margex\ acts on the conformal
field theory of the non-rotating fivebrane (which corresponds to
the limit of $L\rightarrow\infty$ in our parametrization) by
redefining the BRST charge, thus leading to a one-parametric class of
deformations of the BRST cohomology of the model.

Recalling that $\phi\in [0,\pi )$, we can see that the set of
coordinates we have used to describe the
rotating fivebrane covers just one half of the external spacetime.
Obviously, the metric can be continued analytically to $\phi\in[0,2\pi )$;
nevertheless, another way to the analytic continuation
exists.  To show this, let us again start with the Lagrangian \dirsl ,
but now gauge
\eqn\newgau{g_E\rightarrow h_E\; g_E\; h_E^{-1},\qquad g_M\rightarrow
h_M^{\beta}\; g_M\; h_M^{\beta}.}
Repeating the same story as above, we arrive at
\eqn\bacfi{\eqalign{ds_{\rm 5D}^2=&\;\frac{k_M}{4}dr_M^2+\frac{k_E}{4}dr_E^2+
\frac{k_E\cosh ^2\rE (1+L\cosh ^2\rM )}{\sinh ^2\rE +L\cosh ^2\rM}
d\theta ^2\cr
&-\frac{k_M\sinh ^2\rM (L+\sinh ^2\rE )}{\sinh ^2\rE +L\cosh ^2\rM}
dt^2 +\frac{2k_M\beta\cosh ^2\rE \sinh ^2\rM}{\sinh ^2\rE +L\cosh ^2\rM}
dt\; d\theta\cr
&\qquad\qquad +\frac{k_M\cosh ^2\rM\sinh ^2\rE}{\sinh ^2\rE
+L\cosh ^2\rM}d\tilde{t}^2,\cr
B=&\;\frac{k_M\sinh ^2\rE \sinh ^2\rM}{\sinh ^2\rE +L\cosh ^2\rM}dt\wedge
d\tilde{t}-\frac{\beta k_M\cosh ^2\rE \cosh ^2\rM}{\sinh ^2\rE +L\cosh ^2\rM}
d\theta\wedge d\tilde{t} ,\cr
\Phi=&\;\ln (\sinh ^2\rE +L\cosh ^2\rM )+\const\cr}}
where we have used the notation of \bacfive .  Continuing analytically
to the $\sltwo\times SU(2)$ gauged model, we obtain (in the notation of
\fivebrane )
\eqn\fivebra{\eqalign{ds_{\rm 5D}^2=&\;\frac{k_M}{4}dr_M^2+
\frac{k}{4}d\phi ^2+
\frac{k\cos ^2\phih (1+L\cosh ^2\rM )}{\sin ^2\phih +L\cosh ^2\rM}
d\theta ^2\cr
&-\frac{k_M\sinh ^2\rM (L+\sin ^2\phih )}{\sin ^2\phih +L\cosh ^2\rM}
dt^2 -\frac{2k_M\beta\cos ^2\phih \sinh ^2\rM}{\sin ^2\phih +L\cosh ^2\rM}
dt\; d\theta\cr
&\qquad\qquad +\frac{k_M\cosh ^2\rM\sin ^2\phih}{\sin ^2\phih
+L\cosh ^2\rM}d\tilde{t}^2,\cr
B=&\;\frac{k_M\sin ^2\phih \sinh ^2\rM}{\sin ^2\phih +L\cosh ^2\rM}dt\wedge
d\tilde{t}+\frac{\beta k_M\cos ^2\phih \cosh ^2\rM}{\sin ^2\phih +L\cosh ^2
\rM}d\theta\wedge d\tilde{t} ,\cr
\Phi=&\;\ln (\sin ^2\phih +L\cosh ^2\rM )+\const ,\cr}}
which is exactly the analytic continuation of \fivebrane\ from $\phi\in
[0,\pi )$ to $\phi\in [\pi ,2\pi )$.

\newsec{Exact Axisymmetric Stationary Solutions in Four Dimensions}

Now let us return to our attempt at constructing a 4D exact solution
of bosonic string theory by combining two 2D black holes.  The problem
we have arrived at is the violation of Witten's non-anomalousness condition
\ancanc\ by the proposed gauge group action
\gacomp .  In \witan , Witten has shown that even in such anomalous cases, it
is possible to choose a gauged Lagrangian in such a way that the gauge
non-invariant terms don't depend on the WZW group variable.  This leads us
to suspect that a $U(1)\times U(1)$ action on $\sltwo _M\times \sltwo _E$
exists which is still anomalous on each of the
$\sltwo$'s separately, but the gauge non-invariances cancel between
sectors $M$ and $E$.  This is indeed the case, as we are now going to see.

Let us start with \dirsl\ once more, and set $k_M=k_E\equiv k$ for simplicity.
The $U(1)\times U(1)$ group to be gauged acts by
\eqn\uoneuone{\eqalign{U(1)_E:&\qquad g_E\rightarrow h_E\; g_E\; h_E, \qquad
g_M\rightarrow h_M^{\alpha}\; g_M\; h_M^{-\alpha},\cr
U(1)_M:&\qquad g_E\rightarrow h_E^{\alpha}\; g_E\; h_E^{-\alpha}, \qquad
g_M\rightarrow h_M\; g_M\; h_M.\cr}}
It is easy to show that \uoneuone\ satisfies condition \ancanc .  The shift
from \gacomp\ to \uoneuone\ adds new $\alpha$-dependent terms to \lagcoset ,
\eqn\newterms{\eqalign{\CL_{\rm 4D}\rightarrow &\;\CL_{\rm 4D}-
\frac{ik_M}{\pi}\intz \alpha\sinh ^2\rM (A_E\pb t+\AB_E\p t)\cr
&\qquad\qquad\qquad -\frac{2ik_M}{\pi}\intz \alpha ^2\sinh ^2\rM A_E\AB_E\cr
&+\frac{ik_M}{\pi}\intz \alpha (\cosh ^2\rM +\sinh ^2\rM )
(A_E\AB_M-A_M\AB_E),\cr}}
thus modifying the (lowest order) background fields to
\eqn\exfourdi{\eqalign{ds_{\rm 4D}^2=&\;\frac{k}{4}dr_E^2+\frac{k}{4}
dr_M^2+\frac{k\cosh ^2\rM\sinh ^2\rE}{\Delta}
d\theta^2 -\frac{k\sinh ^2\rM \cosh ^2\rE}{\Delta}dt^2,\cr
B=&\;\frac{k\alpha \sinh ^2\rE\sinh ^2\rM}{\Delta}
dt\wedge d\theta ,\cr
\Phi =&\;\ln (\Delta )+\const ,\cr}}
where we have shortened $e^{\Phi}\equiv\Delta$, with
\eqn\delsho{\Delta\equiv \cosh ^2\rM \cosh ^2\rE
-\alpha ^2\sinh ^2\rM \sinh ^2\rE.}
\exfourdi\ is (the lowest order approximation to) the class of exact,
stationary and axisymmetric solutions of four-dimensional string theory
advertised above.%
\foot{This construction can be obviously generalized to
$N=1$ superstrings.  As the cosets will in fact carry $N=2$ supersymmetry,
they will correspond to solutions of $N=1$ superstring theory with $N=1$
supersymmetry in the target.}
To avoid naked singularities, we will restrict ourselves to $|\alpha |<1$.
(Note the singular behavior of the full Lagrangian \newterms\ at the
limiting values of $\alpha$, $\alpha =\pm 1$.)

Now let us analyze shortly the global structure of the solution.  The
surface at $r_M=0$ is an event horizon, which inherits at fixed $t$ the
geometry of the Euclidean 2D black hole:
\eqn\horiz{ds_{\rm horizon}^2=\frac{k}{4}dr_E^2+k\tanh ^2\rE d\theta ^2.}
The structure of the horizon suggests that it might be reasonable to
interpret the solution as a black string.

The external geometry \exfourdi\ can be continued behind the horizon as
follows.  As we have remarked above, the Euler angle parametrization we
have used does not cover the $\sltwo _M$ group manifold completely.  In
another region, the following Euler angles are useful:
\eqn\grpartwo{g_M=\pmatrix{e^{t_L/2}&0\cr 0&e^{-t_L/2}\cr}
\pmatrix{\cos \rM&-\sin\rM\cr\sin\rM&\cos\rM\cr}
\pmatrix{e^{-t_R/2}&0\cr 0&e^{t_R/2}\cr}}
with $r_M\in (-\pi ,\pi )$.  In this parametrization of the Lagrangian,
we are led to
\eqn\intfourdi{\eqalign{ds_{\rm 4D}^2=&\;\frac{k}{4}dr_E^2-\frac{k}{4}
dr_M^2+\frac{k\cos ^2\rM\sinh ^2\rE}{\Delta_{\rm int}}
d\theta^2 +\frac{k\sin ^2\rM \cosh ^2\rE}{\Delta_{\rm int}}dt^2,\cr
B=&\; -\frac{k\alpha \sinh ^2\rE\sin ^2\rM}{\Delta_{\rm int}}
dt\wedge d\theta ,\cr
\Phi =&\;\ln (\Delta_{\rm int})+\const ,\cr}}
now with
\eqn\newdel{\Delta_{\rm int}\equiv \cos ^2\rM \cosh ^2\rE
+\alpha ^2\sin ^2\rM \sinh ^2\rE.}
This describes the geometry of the solution behind the horizon.
Note that $r_M$ has become timelike while $t$ is now spacelike, as might
have been expected.

At $r_M=\pi$ we encounter a singularity.%
\foot{Strictly speaking, this is a future singularity.  The same
analysis can be carried out for the past singularity at $r_M=-\pi$.}
However, an interesting effect
occurs here:  While in the direct product geometry of $\alpha =0$ any
observer behind the horizon must fall into the singularity (which exists
at $r_M=\pi$ for any value of $r_E$), with $\alpha$ nonzero the singularity
is localized at $r_E=0$.  What then happens to the observer at fixed nonzero
$r_E$ with increasing timelike coordinate $r_M$?  It is easy to see that
the internal geometry \intfourdi\ can be continued further to another
region.  This region corresponds to the remaining part of the $\sltwo _M$
group manifold, parametrized by%
\foot{Here $r_M>0$.  The region with $r_M<0$ corresponds to the
continuation through the analogous horizon in the past, and/or to a
`mirror geometry,' analogously as in, say, the Reissner-Nordstr\o m black
hole.}
\eqn\grparthr{g_M=\pmatrix{e^{t_L/2}&0\cr 0&e^{-t_L/2}\cr}
\pmatrix{\sinh \rM&\cosh\rM\cr -\cosh\rM&-\sinh\rM\cr}
\pmatrix{e^{-t_R/2}&0\cr 0&e^{t_R/2}\cr}.}
After crossing the event horizon at $r_M=0$ in \exfourdi , the observer can
avoid the singularity, cross a new, inner horizon at $r_M=\pi$ with
$r_E\neq 0$,
and enter the portion of the universe coming from \grparthr .  At the
inner horizon, $\theta$ becomes timelike, while $r_M$ turns spacelike again.
The former time coordinate $t$ remains spacelike, and plays the role
of an angular variable.  The inner horizon carries the geometry of the dual
Euclidean 2D black hole:
\eqn\innhoriz{ds_{\rm horizon}^2=\frac{k}{4}dr_E^2+\frac{k}{\alpha ^2}
\coth ^2\rE dt ^2.}
Hence, the roles of $t$ and $\theta$ have been completely interchanged
in the region behind the inner horizon, when compared to the geometry
of \exfourdi .  The lowest order background in this region is
\eqn\newfourdi{\eqalign{ds_{\rm 4D}^2=&\;\frac{k}{4}dr_E^2+\frac{k}{4}
dr_M^2-\frac{k\sinh ^2\rM\sinh ^2\rE}{\tilde{\Delta}}
d\theta^2 +\frac{k\cosh ^2\rM \cosh ^2\rE}{\tilde{\Delta}}dt^2,\cr
B=&\; -\frac{k\alpha \sinh ^2\rE\cosh ^2\rM}{\tilde{\Delta}}
dt\wedge d\theta ,\cr
\Phi =&\;\ln (\tilde{\Delta} )+\const ,\cr
\tilde{\Delta}\equiv &\;\alpha ^2\cosh ^2\rM\sinh ^2\rE
-\sinh ^2\rM\cosh ^2\rE .\cr}}
This geometry describes, for $r_M$ close enough to zero, a throat with a
naked singularity.  The throat can be continued through its future
horizon, and the analysis can be repeated infinitely
many times, leading to an infinite strip of geometries and horizons.

Without any computation, the residual (Killing) global symmetry of the
coset is (at least) $U(1)\times U(1)$, as precisely these symmetries
survive the gauging of \uoneuone\ on $\sltwo_M\times\sltwo_E$.
It is worth stressing
that the $U(1)\times U(1)$ is an exact Killing symmetry of the full-fledged,
exact CFT, not just an accidental symmetry of the lowest order background.
In this sense, we are guaranteed to have obtained a class of exact, stationary
and axisymmetric classical solutions of string theory in four dimensions.

The solution we have just constructed is a twisted product of two 2D black
holes.
As the 2D black hole cosets enjoy an interesting property of target duality
\ref\dualpapers{A. Giveon, `Target Space Duality and Stringy Black Holes,'
Berkeley preprint LBL-30671 (April 1991);  R. Dijkgraaf, E. Verlinde and
H. Verlinde, `String Propagation in a Black Hole Geometry,' Princeton
preprint IASSNS-HEP-91/22 \&\ PUPT-1252 (May 1991)}, one might wonder
whether there is an analogy of this stringy symmetry for the 4D cosets.
Actually, for general sigma models with a Killing vector, there is a
duality transformation \ref\buscher{T.H. Buscher, \plb{201}{88}{466}},
given by
\eqn\buschdual{\eqalign{&\hat{G}_{00}=\frac{1}{G_{00}},\qquad
\hat{G}_{0i}=\frac{B_{0i}}{G_{00}},\qquad \hat{G}_{ij}=G_{ij}-
(G_{0i}G_{0j}-B_{0i}B_{0j})/G_{00},\cr
&\qquad\qquad
\hat{B}_{0i}=\frac{G_{0i}}{G_{00}},\qquad \hat{B}_{ij}=B_{ij}+
(G_{0i}B_{0j}-B_{0i}G_{0j})/G_{00},\cr
&\qquad\qquad\qquad\qquad\qquad\hat{\Phi}=\Phi +\ln (-G_{00})\cr}.}
Note that in the case of our 4D solution, we obtain $\hat{B}=0$,
but instead of nonzero values of the antisymmetric tensor field,
nonzero off-diagonal components of the metric tensor occur.
When applied to the region represented by \exfourdi , this duality
transformation would give a metric with a
naked singularity at $r_M=0$, as it is easy to see
that \buschdual\ maps the horizon to a singularity.
To obtain a metric
without naked singularities, it is a better idea to apply \buschdual\
to the region behind the naked singularity in \newfourdi .  Indeed, upon
doing this we obtain a remarkably simple geometry,
\eqn\dualfourdi{\eqalign{\widehat{ds}_{\rm 4D}^2=&\;\frac{k}{4}dr_E^2+
\frac{k}{4}dr_M^2\cr
&\qquad+k\tanh ^2\rE \left[ d\theta^2-2\alpha\; dt\; d\theta -(\tanh ^2\rM
\coth ^2\rE -\alpha ^2)dt^2\right] ,\cr
\hat{B}\equiv&\; 0,\cr
\hat{\Phi}=&\;\ln (\cosh ^2\rE\cosh ^2\rM )+\const .\cr}}
Quite surprisingly, this is a direct product of two 2D black holes!
Indeed, upon changing coordinates to $\Theta =\theta -\alpha t$, $T=t$,
we obtain the standard direct product metric of one Euclidean and
one Minkowskian black hole, parametrized by $r_E,\Theta$ and $r_M,T$
respectively.  How does it come about that the full class
of highly nontrivial spacetimes is dual to a simple, tensor product
structure?  The crucial point is that \buschdual\ assumes a preferred
Killing vector with respect to which the duality transformation is
performed.%
\foot{Compare the recent discussion of duality by Ro\v{c}ek and E. Verlinde
in \ref\rocek{M. Ro\v{c}ek and E. Verlinde, `Duality, Quotients and
Currents,' IAS \&\ Stony Brook preprint IASSNS-HEP-91/68 \&\ ITP-SB-91-53
(October 1991)}.  Note also that there is a similarity
between the duality found above, and the twisting procedure
studied by Sen in \ref\ashsen{A. Sen, `Twisted Black $p$-Brane
Solutions in String Theory,' Tata preprint TIFR/\-TH/\-91-37 (August 1991)}.}
We have tacitly assumed that this Killing vector coincides
with $\p /\p t$.  Nevetheless, it is a peculiarity of our geometry that
there are two commuting Killing vectors in the target, and
the duality transformation now requires a choice of preferred basis
in the space of Killing vectors.  Naively, we can
apply \buschdual\ to any particular choice of the basis,
thus obtaining a class of {\it a priori} different geometries.  The same
situation emerges for the tensor product of two black holes, which
explains the duality observed above.

We have constructed a class of $U(1)\times U(1)$ symmetric solutions
of 4D string theory, and have found indications of a remarkable
duality of the models.  In general relativity, $U(1)
\times U(1)$ symmetric metrics have attracted much interest, and appealing
results have been achieved:  not only many physically
important exact
solutions of Einstein's equations belong to this class, but Einstein's
equations become exactly solvable in this limit by inverse scattering
methods, in particular multi-monopole solutions can be found,
infinite-dimensional solution-generating groups of symmetries exist,
explainable via intimate relations to dimensional reduction to 2D,
to mention at least some of the crucial aspects of axisymmetric stationary
metrics in Einstein gravity.  On the other hand, the 2D black hole cosets,
which serve as basic building blocks for the constructions of our paper,
have been shown recently to enjoy a rich internal structure
\ref\distler{J. Distler and P. Nelson, `New Discrete States of Strings
Near a Black Hole,' PennState \&\ Princeton preprint UPR-0462T \&\ PUPT-1262
(August 1991)}, related in particular to
$W_\infty$ algebras \ref\winf{I. Bakas and E. Kiritsis, `Beyond the
Large N Limit:  Non-linear $W_\infty$ as Symmetry of the SL(2,R)/U(1)
Coset Model,' Berkeley \&\ College Park preprint UCB-PTH-91/44 \&\
LBL-31213 \&\ UMD-PP-92-37 (September 1991)}.
It would indeed be desireable to study possible interplays between
the deep results of general relativity of axisymmetric stationary geometries
on one hand, and string theory
with its extremely rich mathematical structure on the other.
We hope that the exact solutions we have constructed above might serve as a
starting point for further investigation in this direction.

{\bf Acknowledgement.} It is a pleasure to thank Peter Bowcock,
Tohru Eguchi, Jeff Harvey, Elias Kiritsis and Emil Martinec for
valuable discussions.
\listrefs
\bye